\documentclass[11pt, executivepaper]{article}
\usepackage[utf8]{inputenc}
\usepackage[T1]{fontenc}
\usepackage{natbib}
\usepackage{amsmath}
\usepackage{color}
\usepackage{amsfonts}
\usepackage{graphicx}
\usepackage{enumitem}
\usepackage{geometry}
\usepackage{hyperref}
\hypersetup{colorlinks= true, allcolors=blue}
\setcitestyle{aysep={}}

\title{\textbf{No-Go Theorems and the Foundations of Quantum Physics}}

\author{Andrea Oldofredi\thanks{Contact Information: Universit\'e de Lausanne, Section de Philosophie, 1015 Lausanne, Switzerland. E-mail: Andrea.Oldofredi@unil.ch}}

\begin{document}

\maketitle

\begin{abstract}
In the history of quantum physics several no-go theorems have been proved, and many of them have played a central role in the development of the theory, such as Bell's or the Kochen-Specker theorem. A recent paper by F. Laudisa has raised reasonable doubts concerning the strategy followed in proving some of these results, since they rely on the standard framework of quantum mechanics, a theory that presents several ontological problems. The aim of this paper is twofold: on the one hand, I intend to reinforce Laudisa's methodological point by critically discussing Malament's theorem in the context of the philosophical foundation of Quantum Field Theory; secondly, I rehabilitate Gisin's theorem showing that Laudisa's concerns do not apply to it. 
\vspace{4mm}

\emph{Keywords}: No-go Theorems; Quantum Mechanics; Quantum Field Theory; Gisin's Theorem; Malament's Theorem
\end{abstract}
\vspace{3mm}

\begin{center}
\emph{Accepted for publication in Journal for General Philosophy of Science}
\end{center}
\clearpage

\tableofcontents

\section{Introduction}

In the history of quantum physics a considerable number of no-go theorems have been proved, and many of them have played a crucial role for the development of the theory.\footnote{It is interesting to note that physical theories supply salient information on the inherent limitations of knowledge we may have about the world: there are objective matters of fact about it that are not experimentally accessible to us according to specific theoretical frameworks, independently of the current technological resources available. These limitations are derivable from the structure of the given theory at hand, i.e. when axioms and laws of motion are established, and seem to be perfectly suitable examples of no-go results. There are several examples of such limitations in quantum physics, for instance, one may consider that according to QM it is not possible to measure the wave function of an individual system, or that it is not possible to measure the velocity of a particle in Bohmian Mechanics (BM), or that it is impossible to make experiments able to distinguish between BM and QM, or between the mass density and the flash versions of the Ghirardi-Rimini-Weber theory (GRWm and GRWf respectively). Nevertheless, due to the lack of space, in this paper I will not focus on this kind of negative results connected to inherent limitations of knowledge in physical theories.} Typical examples of such results in Quantum Mechanics (QM) are the theorems proved by von Neumann (see \cite{vonNeumann:1955aa}, Chapter 4), J.S. Bell (\cite{Bell:1964aa}) or Kochen and Specker (\cite{Kochen:1967aa}). If one takes Quantum Field Theory (QFT) into account, many no-go theorems deny the possibility to have a particle interpretation of the theory. Among them, well known results are due to Reeh and Schlieder, Haag, Hegerfeldt or Malament. 

Generally, it is fair to say that one should interpret these theorems as formal results that impose boundaries to the theoretical framework in which they are proved (in other words, such theorems tell us what is in principle not possible according to it), establishing limitations about certain mathematical or physical possibilities via a proof by contradiction, suggesting that a specific assumption cannot be true given that one may generate contradictions from it.

Since no-go theorems establish deductively a particular statement from a set of assumptions, it is rather trivial to say that if the deduction of a particular statement rests on wrong, ill-defined, questionable, or circular assumptions, the entire proof is consequently dubious, and the conclusion should not be accepted. Nonetheless, this fact is less trivial considering the current situation within the foundations of quantum mechanics and quantum field theory, where there is no common agreement about the interpretation of these theories. Thus, if we restrict our attention to no-go theorems within the context of quantum physics, it is mandatory to consider carefully both the assumptions on which these theorems rely and the theoretical framework in which they are proposed. 

To this extent, recently \cite{Laudisa2014} has questioned the validity of several no-go results within the foundations of QM. For the sake of clarity, it is important to notice that he does \emph{not} claim that no-go theorems are less important for the conceptual and technical developments of quantum physics than other results proved within this domain. On the contrary, these theorems intend to clarify, imposing particular constraints, the structure of quantum theory; in fact, Laudisa points out correctly that 
\begin{quote}
[...] the significance of proving  ``no-go'' results should consist in clarifying the fundamental structure of the theory, by pointing out a class of basic constraints that the theory itself is supposed to satisfy. (\cite{Laudisa2014}, p. 2)
\end{quote}

Nevertheless, he also underlines that often these theorems rely on dubious assumptions which could be questioned or rejected, since they are formulated within the \emph{standard} or \emph{textbook} framework of QM, a theory which is manifestly affected by conceptual difficulties, and that precludes a clear understanding of the physics at the microscopic level because of the lack of a clear ontology.\footnote{For a general discussion of the conceptual issues in QM the reader may consider \cite{Bell:2004aa}, \cite{Durr:2013aa} and \cite{Bricmont:2016aa}.} According to him, 
\begin{quote}
a sensible way toward such an understanding may be to cast in advance the problems in a clear and well-defined interpretational framework - which in my view means primarily to specify the ontology that quantum theory is supposed to be about - and after to wonder whether problems that seemed worth pursuing still are so in the framework. (\cite{Laudisa2014}, p. 2)
\end{quote}

In this paper I will show another case, following Laudisa's perspective, in which dangerous ontological conclusions against the possibility to maintain a particle interpretation of QFT have been drawn from a (set of) no-go theorem(s) involving questionable metaphysical assumptions, which are based on a ontologically opaque theoretical framework. Thus, the aim of the paper is to reinforce Laudisa's methodological point. However, I am going to begin my discussion objecting against the conclusion he stated concerning Gisin's theorem, showing how to rehabilitate this important result. 
\vspace{2mm}

More precisely, the paper is organized as follows: in Section 2 I will consider a no-go result proposed by N. Gisin, which concerns the impossibility to obtain a relativistically covariant non-local deterministic hidden variable theory that has been already discussed in \cite{Laudisa2014}: I will analyze Laudisa's objections to this argument, and I will show that they are not conclusive. Section 3 is devoted to Malament's theorem; I will argue that its negative conclusions regarding a possible particle interpretation of QFT may be questioned since this latter relies on dubious metaphysical assumptions. The last section contains the conclusions.

\section{On the impossibility of relativistic covariance in non-local deterministic hidden variables quantum theories}

The present case study is an argument contained in \cite{Gisin2011}, which discusses the possibility to build a \emph{covariant} deterministic quantum theory with additional non-local variables. A typical example of such a theory is Bohmian Mechanics (BM). 

This paper provides an impressive result: it is stated that irremediably every non-local deterministic hidden variables theory cannot be made relativistically covariant.\footnote{This result is nowadays under investigation: recently \cite{DGNSZ} and \cite{Horton:2002aa} have argued that there might be possible ways to implement relativistic covariance within the structure of BM.} 
Here, covariant means relativistic time-order invariance w.r.t. theory's predictions, or in other words, it is required that the theory should be invariant under velocity boosts responsible for the time ordering of events in space-time.

Let us introduce the argument in detail. Consider a Bell-type experimental setting, where $A$ and $B$ are two spatiotemporal regions on the same hyperplane, separated by a space-like distance. Suppose that a pair of spin-$1/2$ particles in a singlet state $\psi^-$ are emitted from a source ideally localized between $A$ and $B$. From QM we know that if pairs of particles are prepared in a singlet state, we then expect anti-correlations between the outcomes of measurements performed within the regions $A,B$. Experimenters in $A,B$ choose independently from one another the measurement settings (indicated with $\textbf{a}, \textbf{b}$ respectively). 

Now consider: 
\begin{enumerate}
   \item a class of a reference frames\footnote{This argument assumes a preferred universal reference frame that determines uniquely the temporal order for events in space-time.} $\{F\}$ in which the experimenter in $A$ (following the common jargon, let us call the two experimenters Alice and Bob respectively) is the first to decide her experimental settings, and suppose that she obtains the result $\alpha$, whereas the experimenter in $B$ chooses his settings only after Alice's choice, and obtains $\beta$ as outcome; 
   \item another class of reference frames $\{F'\}$ where $B$ chooses his settings first and obtains the outcome $\beta$, and Alice arranges her experimental settings only after Bob's obtaining $\alpha$. 
\end{enumerate}   
    
The result obtained by Alice in a reference frame $f\in\{F\}$ is 
\begin{align}
\alpha=F_{AB}(\textbf{a}, \lambda, \psi^-) \nonumber
\end{align} 

\noindent{which} is a function of the settings $\textbf{a}$, the non-local hidden variables $\lambda$ and the particle state $\psi^-$ (the label of the function, $F_{AB}$, reminds us that Alice chooses first her experimental setting and Bob chooses after her). In $f$, Bob obtains
\begin{align}
\beta=S_{AB}(\textbf{a}, \textbf{b}, \lambda, \psi^-) \nonumber
\end{align} 

\noindent{which} depends (as expected) also on Alice's settings: ``this is the sense in which the variable $\lambda$ together with the function $F_{AB}$ and $S_{AB}$ form a non-local model''. (\cite{Gisin2011})

Let's consider instead a reference frame $f'$ which is a member of $\{F'\}$. Here, Bob's result is given by
\begin{align}
\beta=F_{BA}(\textbf{b}, \lambda, \psi^-) 
\end{align}

\noindent{and} by applying the same reasoning we consequently get
\begin{align}
\alpha=S_{BA}(\textbf{a}, \textbf{b}, \lambda, \psi^-). 
\end{align} 

A this point, in order to derive Gisin's conclusion we need to introduce the \emph{covariance} condition. In a covariant non-local model Alice and Bob's results should be \emph{independent} of the reference frame, and this follows from the principle of relativity. Therefore, it follows that
\begin{align}
\alpha=F_{AB}(\textbf{a}, \lambda, \psi^-)=S_{BA}(\textbf{a}, \textbf{b}, \lambda, \psi^-)
\end{align} 

\noindent{and} similarly 
\begin{align}
\beta=F_{BA}(\textbf{b}, \lambda, \psi^-)=S_{AB}(\textbf{a}, \textbf{b}, \lambda, \psi^-).
\end{align} 

However, the astonishing conclusion to draw from equations (1)-(2) is that they define a \emph{local} model in Bell's sense, since from (3) one deduce that the function $S_{BA}$ is independent of $\textbf{b}$, and the same reasoning applies to (4). But, it is well known that local models cannot reproduce the predictions of QM according to Bell's theorem. Hence, any covariant deterministic non-local hidden variables model is \emph{equivalent} to a local model, and consequently it is inconsistent with quantum predictions. 

In conclusion, following \cite{Gisin2011} there exists no covariant non-local deterministic hidden variables model. 

Briefly stated, the strategy followed has been to show that every covariant non-local deterministic hidden variable model turns out to be a \emph{local} model; therefore, by a simple application of Bell's theorem, we obtain that it is impossible to have a non-local deterministic hidden variables model that satisfies relativistic covariance. 

After having introduced Gisin's no-go theorem, it is interesting to understand the reasoning followed by Laudisa to argue against this type of argument. He raises the following objections against Gisin's model:

\begin{enumerate}
   \item Considering the equations (3) - (4) Laudisa points out that $\lambda$ has a non-local character that depends on the first choice of the experimental settings: in the case of (3) the hidden parameter is non-local only for Bob, whose result $\beta$ depends on ($\lambda + \textbf{a}$), vice versa, in (4) it is non-local only for Alice. It seems that non-locality manifests itself only for a \emph{single party at a time}. Laudisa stresses that applying the covariance condition implies to impose an equality between $S_{BA}(\textbf{a}, \textbf{b}, \lambda, \psi^-)$ and $F_{AB}(\textbf{a}, \lambda, \psi^-)$, where the latter is a local model from the beginning. Thus, by assuming a local hidden parameter $\lambda$ at the beginning, the conclusion seems to become circular.
   \item The second objection makes use of a possibility which is not considered by Gisin. What happens if we consider a reference frame in which Alice and Bob's measurements are \emph{simultaneous}? The correct question is: in the case in which the measurements would be performed simultaneously, would the model turn out to be local or non-local? Laudisa claims that the results $\alpha,\beta$ would be defined by the functions $\alpha=SIM_{AB}(\textbf{a}, \textbf{b}, \lambda, \psi^-)$ and $\beta=SIM_{BA}(\textbf{b}, \textbf{a}, \lambda, \psi^-)$ respectively. From these latter formulations of $\alpha$ and $\beta$, it is easy to derive the logical consequence from the covariance requirement:
   \begin{align*}
   \alpha=SIM_{AB}(\textbf{a}, \textbf{b}, \lambda, \psi^-)=F_{AB}(\textbf{a}, \lambda, \psi^-)=S_{BA}(\textbf{a}, \textbf{b}, \lambda, \psi^-)
   \end{align*}
Thus, the model is local or non-local according to a specific choice of the reference frame. Such a disappointing feature is inherited from the fact that $\lambda$ works as if it were a local parameter. Thus, it is not clear how covariance could turn a non-local model into a local one. 
\end{enumerate} 

These points are undoubtedly interesting and deserve to be discussed further. The first thing to highlight is that the basic idea behind the definition of $\alpha$ is that if Alice is the first to choose her experimental settings, this choice \emph{cannot be dependent on Bob's}: he will choose his settings only after Alice (this fact follows from the definition of the experiment under discussion). If we ascribe a non-local character to $\lambda$, e.g. we consider it as a non-local parameter, it is easy to see that from the instant of time in which Alice chooses her settings we have a simultaneous non-local effect that will affect the configurations of the entities, \emph{whichever they are}, that compose Bob's experimental set up. For instance, if the model implements a particle ontology exactly as in BM, we would observe that a slight modification of the particles' configuration in $A$ (Alice's choice of settings), at whatever instant of time $t_0$, would instantaneously influence every other particle of the configuration (including Bob's experimental settings). This is what is meant by Gisin when he claims that Bob's choice is dependent even on Alice's. Therefore, the definition of $\alpha=F_{AB}(\textbf{a}, \lambda, \psi^-)$ is logically and physically sound. 

Gisin constructed his argument using two different sources of non-local influences which come from Alice's and Bob's choices respectively, depending on the reference frame we decide to choose, $f$ or $f'$. This move, however, is perfectly consistent and it would be absurd to require the opposite, i.e. that Alice's result would be dependent on a choice of settings that Bob has not already made, and vice versa. Furthermore, we should underline that Gisin is considering the most general case conceivable, since he does not pose any condition on the type of variables $\lambda$ should represent. This generality is a virtue of the model. 

I would also like to stress another fact: so far, we have been considering two different counterfactual situations, one in which Alice is the first experimenter making the choice of the settings, and one in which Bob chooses first. These two different physical situations imply two different consequences. In the first case, we expect that Bob's choice will depend upon Alice's, so the non-local effect propagates from $A$ to $B$; in the reverse situation instead, Alice's choice will depend upon Bob's, so that the non-local effect propagates from $B$ to $A$. This propagation is instantaneous given the non-locality of the model considered by Gisin: the source of non-locality comes from the experimenter that first changes the physical situation. Bob's choice is instantaneously affected by Alice's: she has to choose her settings, and in doing so she modifies the \emph{entire} experimental situation. This reasoning applies equally to Bob's choice. Thus, again, Gisin's definitions of $\alpha$ and $\beta$ are logically sound. Furthermore, equations (3) and (4) are only consequences of the application of the principe of relativity. Thus, every step of Gisin argument is logically correct, not circular.

Nevertheless, on this basis, it could be said that the unnatural result of the application of the covariance condition that Laudisa exhibits is exactly the price to pay for an absolute and unique time for the configuration of the hidden parameters: the requirements of relativistic covariance are \emph{not} satisfied and, \emph{as a consequence}, we are confronted with a very unphysical situation, that is to turn a non-local model into a local one. 

I would like to conclude this section with a final comment on some consequences of Gisin's result for the foundations of BM and QM. Bohmian mechanics is the clearest instance of a non-local, deterministic, hidden variables quantum theory where the additional parameters are particles positions which follow definite trajectories in physical space (see for details \cite{Durr:2009fk} and \cite{Durr:2013aa}). It is obvious what kind of information about this theory is provided by Gisin's theorem: if this latter is correct, then a theory such as BM cannot be made relativistically covariant. But this would be instructive for the foundations not only of BM, but of quantum theory in general: such a discussion could be helpful to understand the possibility or the \emph{impossibility} to make a certain class of quantum theories genuinely relativistically covariant. In other words, this could be a compelling argument against the possibility to combine an ontologically well-defined non-local, deterministic, hidden variable quantum theory with the theory of Special Relativity (SR) \emph{without} any substantial modification of the requirement of Lorentz Invariance. 

Secondly, Gisin's no-go theorem can be a helpful result suggesting a possible right track regarding future research in quantum theory. It might force philosophers and physicists (i) to understand what are the difficulties to unify SR with a quantum theory with a clear primitive ontology on the one hand, i.e. immune from the usual quantum puzzles, and (ii) to find alternative strategies to combine the two on the other (consider for instance \cite{DGNSZ}, \cite{Hiley2010} and \cite{Dewdney2001}).  

In conclusion, given that the assumptions and the definitions of Gisin's theorem are logically and physically sound, it is not affected by ill-defined notions or circular moves. Furthermore, given the consequences I have mentioned above, it seems that a rehabilitation of this result should be a welcome result. 

\section{Malament's no-go theorem: Localizability of particles in QFT}

In this section I will consider an influential no-go theorem proved by D. Malament which lies at the heart of the philosophical foundations of QFT, and that questions one of the main properties of the particle concept: localizability. The statement of this theorem has had remarkable resonance, generalizations and improvements, and it has now become the standard position concerning the ontology of QFT in philosophy of physics. Nevertheless, in this case Laudisa's concerns are more than justified, since Malament's argument against a particle interpretation of quantum field theory is formulated within the \emph{standard framework} of relativistic QM, which by construction inherits the ontological problems of ordinary (non-relativistic) quantum mechanics, as we shall see in the remainder of this section. 
\vspace{3mm}

What this theorem primarily aims to show is that the requirements of relativity theory and quantum mechanics cannot be reconciled in a relativistic quantum theory of localizable particles. Let us see, then, how localizability has been questioned by Malament's theorem\footnote{In \cite{Halvorson2002} generalizations of Malament's theorem are provided, but for the purposes of the paper is sufficient to consider the original result; in the present discussion I heavily rely on their exposition of Malament's argument.}, and whether there are convincing reasons to give up the particle concept and to abandon a particle interpretation of QFT.

The main claim of \cite{Malament1996} is that in an attempt to reconcile the axioms of QM with special relativity one is led to a field ontology. The logic of the argument could be summarized as follows: if the probability to detect a particle in space-time is constantly zero, then there is no possibility to measure nor to detect it. If a particle is not detected, then it is not localized in space-time. But it is an obvious fact that if an object is a particle, it must be a localized object (or better it must be \emph{localizable}, detectable by particle detectors). Hence, the conclusion is straightforward: if it is not possible to localize particles in space-time, simply there are no particles in it. 

The arena in which the argument takes place is Minkowski space-time $M$, and let $S$ be a family of spacelike hyperplanes covering $M$. 
\vspace{3mm}

\textbf{Definition 1:} \emph{Spatial set}. A spatial set is a bounded open set $\Delta\in{S_i}$ in one of the particular hyperplanes $S_i\in{M}$.
\vspace{3mm}

No specification about the size of $\Delta$ is made, thus, the definition is completely general.
We express the fact that $\Delta$ and $\Delta'$ are bounded spatial regions of $M$ separated by a space-like interval writing $\Delta\cap\Delta'=\emptyset$. 
Now it is necessary to define a \emph{localization system}.   
\vspace{3mm}

\textbf{Definition 2:} \emph{Localization system}. A localization system is a triple $(\mathcal{H}, \Delta\mapsto{E_{\Delta}}, \textbf{a}\mapsto{U(\textbf{a})})$ over $M$. 
Let us give a closer look to its elements:
\begin{enumerate}[label=(\roman*)]
   \item the pure states in QM are described as rays in Hilbert space $\mathcal{H}$, the state space of quantum systems;  
   \item the mapping $\Delta\mapsto{E_{\Delta}}$ goes from bounded open subsets of some $S$ to projections in $\mathcal{H}$. It is an assignment to every spatial set of $M$ of a family of projector position operators $E$. Here $E_{\Delta}$ represents the proposition: \emph{a certain position measurement finds with certainty a particle within $\Delta$}. Thus, a detection of a particle in any given $\Delta$ is represented by projection operator $E_\Delta$ on $\mathcal{H}$;
   \item the mapping $\textbf{a}\mapsto{U(\textbf{a})}$ is a strongly continuous representation of the translation group of $M$ in $\mathcal{H}$.\end{enumerate}
It is important to underline that the \emph{local} operation of particle detection within this context is represented by a self-adjoint projection operator $E$ associated with spatial region $\Delta$, meaning that within that region a detector has been placed. Being a projection operator, $E$ will have two eigenvalues 1 and 0, representing respectively (i) that a particle is localized in some $\Delta$ with certainty (the detector clicks), and (ii) that no particle has been detected within $\Delta$ (the detector does not click). 

Once the localization system is defined, Malament imposes four conditions on it in order to state the theorem (in what follows I use the formulation of these conditions provided in \cite{Halvorson2002}, sec. 2):
\begin{enumerate}
   \item \emph{Translation Covariance}: 
   For any $\Delta$ and for any translation in $M$
\begin{align}   
U(\textbf{a})E_{\Delta}U(\textbf{a})^*=E_{\Delta+\textbf{a}}. \nonumber
\end{align} 

   \item \emph{Energy Bounded Below}:
   For any timelike translation \textbf{a} of $M$, the generator $H(\textbf{a})$ of the one-parameter group $\{U(t\textbf{a}): t\in{R}\}$ has a spectrum bounded from below.
  
   \item \emph{Localizability}:
   If $\Delta$ and $\Delta'$ are disjoint subsets of a hyperplane $S_i$ of $M$ ($\Delta\cap\Delta'=\emptyset$), then
\begin{align}
E_{\Delta}E_{\Delta'}=0. \nonumber
\end{align}
   \item \emph{Microcausality}:  
   If $\Delta$ and $\Delta'$ are disjoint subsets of a hyperplane $S$ of $M$ ($\Delta\cap\Delta'=\emptyset$), and if the distance between $\Delta$ and $\Delta'$ is non-zero, then for any time-like translation \textbf{a}, there is an $\epsilon>0$ such that $[E_{\Delta},E_{\Delta'+t\textbf{a}}]=0$, whenever $0\leq{t}\leq\epsilon$.     
\end{enumerate}

Now we are ready to state the theorem:
\vspace{3mm}

\textbf{Theorem (Malament)}
Let be $(\mathcal{H}, \Delta\mapsto{E_{\Delta}}, \textbf{a}\mapsto{U(\textbf{a})})$ a localization system in $M$ satisfying conditions (1) - (4), then
\begin{align}
E_{\Delta}=0, \forall\Delta\in{M}.
\end{align}

Now let us introduce the motivations for each condition Malament imposes on the localization system. 
\emph{Translation covariance} tells us that if we spatially translate a given particle by a certain vector $\textbf{a}$, then its original wave function $\psi(x)$ transforms into $\psi(x)_\textbf{a}$. This condition is a reasonable constraint since its physical meaning is that the statistics of a given measurement must not change with spatial translations: $\langle\psi,E_{\Delta}\psi\rangle=\langle\psi_\textbf{a},E_{\Delta+\textbf{a}}\psi_\textbf{a}\rangle$. The second condition is a constraint imposed on the Hamiltonian: it must be bounded from below. This means that this operator, the energy operator, has a ground energy state. This implies that it cannot be possible to extract arbitrarily (even infinite) amount of energy from a given particle.

\emph{Localizability} is the first condition involving the particle concept properly, indeed it states that a particle cannot be detected in two different places simultaneously. 

If projective operators are associated with the operation of particle detection and if we consider, exactly as in the case of Malament's theorem, a single particle case, then it is a natural demand to require that the system is not detectable in two space-like separated regions at the same time. This is mathematically stated by saying that the product of the operators $E_{\Delta},E_{\Delta'}$ vanishes. Consider $\Delta, \Delta'$, two space-like separated regions on a particular $S_i$ in which it is placed a particle detector. Each detector will click (meaning that a particle has been detected) with probability $\langle\psi|E_{\Delta}\psi\rangle$ for some initial wave function $\psi\in\mathcal{H}$. Being $E_{\Delta}$ a projection operator, it has only two possible eigenvalues 0 and 1. Therefore, the eigenstates corresponding to situations in which the detector will click with certainty are mathematically represented by $\langle\psi|E_{\Delta}\psi\rangle=1$. In these cases, if the detector in $\Delta$ clicks with certainty and if we consider a one-particle state, then the probability that the detector placed in $\Delta'$ clicks must be $\langle\psi|E_{\Delta'}\psi\rangle=0$. $E_{\Delta}$ and $E_{\Delta'}$ are mutually orthogonal. All this entails that $E_{\Delta}E_{\Delta'}=0=(E_{\Delta}E_{\Delta'})^{\dagger}=E_{\Delta'}E_{\Delta}$ must be valid. 

This is a minimal condition one should require, and it represents a basic fact implied by the notion of particle's position: given that for every particle its position is uniquely defined at every instant of time, it is not possible that a single particle can have two different positions in space (in the same $S_i$) at the same time $t$. Hence, \emph{Localizability} seems a necessary condition to define the particle notion, and it seems totally reasonable to assume the orthogonality of $E_{\Delta}$ and $E_{\Delta'}$. This condition is particularly weak since it does not say anything regarding any finite limit of the velocity of light. 

The last condition is \emph{Microcausality}, the only genuinely relativistic condition imposed on the localization system and it could be derived from the fact that we consider two space-like separated regions $\Delta\cap\Delta'=\emptyset$. Being space-like separated, $\Delta$ and $\Delta'$ have disjoint neighborhoods. From this fact it follows that if operators corresponding to local measurements are assigned to these space-like separated regions, then it is a natural demand to require that these measurements do not mutually interfere. In other words, this condition states that measurements performed within a given space-time region $\Delta$ cannot influence the statistics of measurements performed within another spatial region $\Delta'$, where $\Delta$ and $\Delta'$ are space-like separated. Furthermore, it imposes that the measurement results should not depend on the temporal order of the experiments (the results must be independent of the choice of the Lorentz frame of reference).
The consequence of the failure of \emph{Microcausality} is that superluminal signals are admitted within the theory, but this fact would contradict the axioms of SR, according to which the velocity of light $c$ is the upper bound on the speed at which physical signals can be propagated. In this specific context its failure would imply that measurements performed within $\Delta$ would influence the statistics of measurements performed within $\Delta'$, where they are space-like separated regions. Nonetheless, the local behavior of quantum objects in relativistic QFT is a fundamental requirement, because the theory's structure has been designed to be in agreement with the axioms of SR.\footnote{It is interesting to note that \cite{Malament1996}, p. 2 carefully analyzes the \emph{costs} implied by a QFT with a particle ontology, which is an \emph{unacceptable} non-local act-outcome correlation:
\begin{quote}
I want to use the theorem to argue that in attempting to do so (i.e. hold on to a particle theory), one commits oneself to the view that the \emph{act} of performing a particle detection experiment here can statistically influence the \emph{outcome} of such experiment there, where ``here'' and ``there'' are space like related. [...] I have always taken for granted that relativity theory rules out ``act-outcome'' correlations across space like intervals. For this reason, it seems to me that the result does bear its intended weight as a ``no-go theorem''; it \emph{does} show that there is no acceptable middle ground between ordinary, non-relativistic (particle) mechanics and relativistic quantum field theory. 
\end{quote}
It is clear that QFT here is explicitly intended to be a combination of QM and SR: non-local correlations violate relativistic causality, then if a particle theory implies such non-locality it must be rejected.}$^,$\footnote{I have to thank C. Beck for his helpful and extensive comments on this topic.} 

What this theorem does show is that there cannot be a particle mechanics that respects both the axioms of quantum mechanics and those of special relativity, properly combined by the conditions (1) - (4). The theorem shows that if it is not possible to localize particles in space, then a relativistic quantum mechanical theory should not be a particle theory. The particle notion implies by definition the property of localization in space, then if particles are non localizable objects, we have reasonable arguments against a particle interpretation of the theory. After all, it would seem quite strange to have particles as fundamental entities of a theory which predicts that every position measurement is invariably zero for every region of space-time. 

In conclusion, this theorem shows the impossibility to obtain a physically acceptable notion of localizability of a quantum particle in QFT. Thus, Malament's theorem takes the form of a no-go result for a particle interpretation in relativistic quantum theory.
\vspace{2mm}

Malament's theorem is technically exemplary; nonetheless, its ontological consequences are questionable. At a first sight, the metaphysical lesson that is possible to infer from Malament's argument is that all \emph{talk about particles} should be understood as \emph{talk about fields} in the sense that, since a particle ontology seems to be totally inadequate in the context of quantum field theory, a field ontology should be instead the right candidate. Here a bivalence principle is in action: given that only two possibilities are valuable for the ontology of the theory (either particles or fields), if a mechanics of particles is excluded, the only possibility is the concept of field. This is the positive content of Malament's theorem. Nonetheless, even considering a field ontology does not immediately solve the interpretational problems of QFT since \cite{Baker2009} shows that in the context of the \emph{algebraic} approach to QFT the arguments against the possibility of a particle ontology (consider especially the work done by D. Fraser in \cite{Fraser2006}, \cite{Fraser2008}) have exactly the same consequences against a field interpretation of the theory. Thus, also a field ontology seems to be inadequate to represent the basic entities of QFT. Baker argues then in favor of an ontology based on the notion of local algebras, following Halvorson and Clifton, as it appears clearly considering sec. 7 of \cite{Halvorson2002}. This is another different way to answer to the ontological question about the nature of the fundamental objects in QFT, but this line of reasoning is notoriously affected by the problems emphasized in several places by Bell (collected in \cite{Bell:2004aa}), thus, it seems difficult to take seriously into account such kind of entities as fundamental objects in one's ontology.\footnote{For lack of space I cannot recall here all the unwelcome implications of the identification of local observables with local beables; the reader should refer to \cite{Durr:2004c} for a technical discussion.}

It is crucial to underline that in all these arguments the \emph{local beables} are identified with \emph{local observables}. Indeed, this theorem identifies the position of a particle with an effective operation, namely, a position measurement in space-time. Nevertheless, one may legitimately ask whether we should be convinced by the ontological conclusions of this argument, or if this theorem is concerned with ontologically \emph{secondary} notions. 

In the remainder of the section I will argue in favor of the second option, taking into account several arguments elaborated within the primitive ontology programme:\footnote{The following arguments apply even to the theorems contained in \cite{Halvorson2002}.}
\begin{enumerate}
   \item There are successful theoretical frameworks as Bohmian Mechanics, Nelsionian mechanics or the extensions of BM to quantum field theory (BQFT), where the meaning of the position operator is \emph{not} that of an observable. Its physical meaning is given by the fact that a certain particle \emph{is} in a certain position $x\in\mathbb{R}^3$ in physical space at every time $t$, since according to these theories particles always have definite positions and follow trajectories in space. Taking into account, for example, the axioms of BM and BQFTs, there is no room for physically ill-defined notions as measurement or observer: these are secondary notions which ontologically depend on the local beables. 
   
More specifically, in the context of BM operators have the role to \emph{connect} vectors in $\mathcal{H}$ to points in physical space $\mathbb{R}^3$. Thus, BM acknowledges the difference between the actual position the particles have in space and what particle detector measures, and in the following lines I will discuss an example showing that these two notions should not be considered equivalent. Furthermore, \emph{contra} Malament's claim, it is significant for our discussion to consider the literature concerning the BQFTs (e.g. \cite{Durr:2004aa}, \cite{Durr:2005aa} and \cite{Colin:2007aa} or \cite{Struyve:2010aa} for an overview). Many of the argument contained in these works show in fact that there are quantum field theories able to implement a particle ontology, which are also mathematically consistent and empirically adequate. Moreover, to prevent the easy objection that BQFTs are not interesting theories because of their not being relativistic, one may consider two simple facts: (i) even the standard model of particle physics is not a genuine relativistic theory since it implements cut-offs in order to have well-defined and well-behaved Hamiltonians, and (ii) the algebraic approach to QFT is \emph{not} empirically adequate since it does not reproduce any model with (realistic) interactions in four dimensions. In addition to this, the problem to define a consistent combination between SR and QM applies to every possible quantum theory currently at our disposal. To this regard, it is important to stress that \emph{every} Bohmian theory maintains by construction a crucial feature of SR: Bohmian particles do not travel faster than the velocity of light and the quantum equilibrium hypothesis ensures that superluminal signaling is completely avoided. Thus, although BM and BQFTs are non-local theories (violating the last condition imposed by Malament), they do not imply any faster than light signaling, and are totally physically adequate.
   \item In BM it is also possible to introduce the difference between \emph{ideal} and \emph{measured} position operators: in the first case the operators refer to the real position the particles have in physical space, in the second case the operators refer to what detectors measure. There are cases in which position measurements \emph{do not} reveal the position of the particle. An example is discussed in \cite{Durr:2004c} (sec. 7.5) and it will be reported for its simplicity and usefulness. D\"urr, Goldstein and Zangh\`i (DGZ) found an appropriate case in which a \emph{position measurement does not measure the position of the particle.} It is a trivial fact that in BM the particles' position is a measurable quantity and to know the position of a particle a measurement of the position operator is needed. However the converse statement is less obvious, since there exist measurements of position operator that do not reveal the exact position of a Bohmian particle. DGZ consider a quantum harmonic oscillator in two dimensions with Hamiltonian 
\begin{align*}
H=-\frac{\hbar^2}{2m}\Bigg{(}\frac{\partial^2}{\partial{x}^2}+\frac{\partial^2}{\partial{y}^2}\Bigg{)}+\frac{\omega^2m}{2}(x^2+y^2) 
\end{align*}
\noindent{and} note that the evolution of the wave function $\psi_t$ is periodic (``except for an irrelevant phase factor'') with period $\tau=2\pi/\omega$. They claim (\cite{Durr:2004c}, p. 1016):
\vspace{2mm}

\begin{quote}
[t]he Bohm motion of the particle, however, needs not have period $\tau$. For example, the $(n=1,m=1)-$state, which in polar coordinates is of the form
\begin{align*}
\psi_t(r,\phi)=\frac{m\omega}{\hbar\sqrt{\pi}}re^{-\frac{m\omega}{2\hbar}r^2}e^{i\phi}e^{-i\frac{3}{2}\omega{t}},
\end{align*}
\noindent{generates} a circular motion of the particle around the origin with angular velocity $\hbar/(mr^2)$, and hence the periodicity depending upon the initial position of the particle - the closer to the origin, the faster the rotation. Thus, in general,
 \begin{align*}
X_{\tau}\neq{X_0}.
\end{align*} 
\end{quote}

What is important, nonetheless, is that although the equation $X_{\tau}={X_0}$ is not generally valid,  $X_{\tau}$ and $X_0$ are random variables \emph{identically} distributed according to the usual $|\psi|^2$-distribution, hence $|\psi_{\tau}|^2=|\psi_0|^2$. 

In order to support the above mentioned thesis according to which not every measurement of the position operator is a genuine position measurement, DGZ propose the following argument: consider two experiments which begin at the same time $t$: $E_1$ and $E_2$, where the former is a measurement of the \emph{initial} position $X_0$ of the Bohmian particle - the position operator - and the latter is a measurement of the position $X_{\tau}$, meaning that it is the position at time $\tau$ to be observed. From the fact that $X_0$ and $X_{\tau}$ are identically distributed, it follows that 
for all $\psi$ the result of $E_2$, which measure the position of the particle at time $\tau$, has the \emph{same distribution} of the experiment measuring $X_0$, $E_1$: therefore, also $E_2$ counts as a measurement of the position operator. But $E_2$ is \emph{not} a measurement of $X_0$, i.e. of the initial position of the particle, since in general at time $\tau$ "does not in general agree with the initial position" (\cite{Durr:2004c}, p. 1017). Hence, they conclude: \emph{a measurement of the position operator is not necessarily a genuine measurement of [the] position [of a Bohmian particle]} (parantheses added). The reader may refer to \cite{Durr:2004c}, \cite{Daumer1996} and \cite{Bell:2004aa} for arguments against the identification between operators and genuine properties of quantum systems.

   \item Particle detectors are necessarily localized in space. Suppose a particular detector is placed within a region $\Delta\in\mathbb{R}^3$. Now, if we consider a detector to be a yes-no experiment, with $Z=1$ if the detector clicks and $Z=0$ if it does not, for any outcome $Z=1$ there is a particle in $\Delta$ (assuming that there is a perfect correlation between the result $Z=1$ and the particle being in $\Delta$), then, the joint wave function of the particle and the apparatus will be $|\psi|^2-$distributed; moreover, the part of $\psi$ representing the detector outcome ($Z=1$) would be supported in the set $\Delta$ which is compact. But this would be in contradiction with the fact that wave function cannot have compact support in $\mathcal{H_+}$. This would be another argument against a particle interpretation of the theory. Nevertheless, from that it does not follow that the particle does not have a position in space.\footnote{This idea is originally contained \cite{Tumulka2014}. I owe this point to R. Tumulka, W. Myrvold and C. Beck.} 
Indeed, considering the one-particle Bohm-Dirac model introduced in \cite{Bohm:1953aa} one notes that there is no contradiction between the possibility of a particle ontology on the one hand, and the fact that wave function cannot have arbitrarily narrow supports in $\mathcal{H_+}$ on the other. Particles have always defined positions in space, albeit wave function are not arbitrarily narrow. The spreading of the wave function is a well-known fact from ordinary QM, and it implies an uncertainty which is \emph{epistemic} and corresponds to the accuracy with which we can know the actual position of the particle. No uncertainty concerning the ontology of the theory is at stake here.\footnote{For the $N$-particle case Bohm introduced the Dirac sea hypothesis, for recent developments see \cite{Colin:2007aa}.} 
   \item \cite{Barrett2002} correctly claims that metaphysical considerations do a lot of substantial work in order to construct a meaningful physical theory. In particular, he correctly claims that the explanation of definite measurement outcomes, as the one considered in Malament's theorem, must be derived from the basic entities of QM and relativistic quantum mechanics. These outcomes, one would say, must be explained in terms of the ontology of the theory, in perfect agreement with Bell's message. Unfortunately, QM is affected by the measurement problem and, as a matter of fact, it is completely inherited by quantum field theory (see especially \cite{Barrett2002} and \cite{Barrett:2014aa}), to this regard Barrett writes:
\begin{quote}
The point here is just that in quantum mechanics one's metaphysical commitments must be sensitive to how one goes about solving the measurement problem. Indeed, it seems to me that no metaphysics for relativistic quantum field theory can be considered satisfactory unless determinate measurement records somehow show up in one's description of the world. Put another way, one must have a solution to the quantum measurement problem before one can trust any specific interpretation of relativistic quantum field theory. (\cite{Barrett2002}, 167.)
\end{quote} 
More specifically on this point, relativistic quantum mechanics does not solve the measurement problem, for it does not provide any physical description in terms of its fundamental objects which would be able to explain the experimental results one obtains in performing measurements on a given system. 

Leaving aside these crucial problems of relativistic quantum mechanics, which \emph{per se} should be a symptom of the conceptual inadequacy of the theory, Barrett interestingly discusses the notion of detectability, which is central in Malament's theorem. In order to be detectable, a measurement record should be about something \emph{which it is effectively possible to find in a finite region of space $\Delta$}, and in quantum mechanics this property of being detectable in such finite region $\Delta$ is mathematically represented by a projection operator defined on it.
The crucial point in Malament's argument is that there is no such \emph{record-detection} operator, as Barrett calls it. To this regard
\begin{quote} 
[a] natural reaction would be to deny the assumption that a detectable record token is a detectable entity that occupies a finite spatial region and insist that in relativistic quantum field theory, as one would expect, all
determinate record tokens are represented in the determinate configuration of some unbounded field. After all, this is presumably how records would have to be represented in any field theory. (\cite{Barrett2002}, p. 174.)
\end{quote}
To see the problem from another, clearer perspective, let me say that surely it is possible to claim (as Malament seems to do) that measurement outcomes could be meaningfully represented by unbounded field configuration, but, as Barrett correctly points out, we are confronted with two problems. The first is the classic measurement problem, which persists unaltered from QM, and the second comes from the scientific practice: the measurements we perform seem to have locations, we use their spatial properties in order to individuate, read and interpret the results. To state something similar to what Malament claimed with his argument seems to go against the actual scientific practice and the experimental evidence which empirically supports relativistic quantum mechanics: it seems that  experimental evidence seems to support the idea of determinate trajectories of (supposed) fundamental particles; it is then evident that if these objects do not exist, it would be extremely hard to find an explanation for such an evidence.
The puzzle is the following: Malament's theorem and its generalization do not prevent the possibility for a particle to have a position, but they rule out the possibility for such particles to have \emph{detectable} positions; however, this claim goes against the actual scientific practice, since detectable positions of some objects are just the sort of experimental evidence we have. More precisely, these detectable positions are just the only things we obtain while measuring something. 
Finally, 
\begin{quote}
if detectable spatio-temporal objects are incompatible with relativistic quantum mechanics, then the challenge is to explain why it seems that we and those physical objects to which we have the most direct epistemic access (our measurement records) are just such objects [footnote deleted]. As far as I can tell, it is possible that all observers and their records are somehow represented in field configurations; it is just unclear how the making, finding, and reading of such records is supposed to work in relativistic quantum field theory. (\cite{Barrett2002}, p.176.) 
\end{quote}
This problem is substantially left without any answer from Malament's claim and it could be interpreted as a sort of final verdict about its plausibility.\footnote{Another argument against Malament's claim and its generalization focused on the current scientific practice is contained in \cite{MacKinnon2008}.}
\end{enumerate}

With these remarks I argued that Malament's theorem is not a result with a substantial ontological import, thus, Laudisa's criticisms are absolutely appropriate in this context. 

The main questionable assumptions concern the identification of position operators with operations of particle detection. I have shown that in ontologically clear frameworks, where there is no flattening of the notion of local beables on that of local observable, it is still possible to propose a (mathematically and physically consistent) QFTs supporting a particle ontology. 

However, as already underlined, the problem to find a consistent relativistic QFT remains open: (i) standard relativistic QM inherits by construction the conceptual issues of ordinary QM (furthermore, even standard model is not genuine relativistic), (ii) the algebraic approach to QFT is not empirically adequate and relies on debatable metaphysical assumptions, and (iii) BQFTs are not yet relativistic theories. This current situation, nonetheless, should be a stimulus for further research.
\vspace{2mm}

It is instructive to conclude this section with a quotation on the secondary significance of many concepts that unfortunately appear in the axioms of both QM and QFT and consequently even in Malament's theorem:
\begin{quote}
The concept of ``observable'' lends itself to very precise mathematics when identified with ``self-adjoint operator''. But physically, it is a rather wooly concept. It is not easy to identify precisely which physical processes are to be given the status of ``observations'' and which are to be relegated to the limbo between one observation and another. So it could be hoped that some increase in precision might be possible by concentration on the \emph{be}ables, which can be described in ``classical terms'', because they are there. [...] The beables must include the settings of switches and knobs on experimental equipment, the currents in coils, and the reading of instruments. ``Observables'' must be \emph{made}, somehow, out of beables. The theory of local beables should contain, and give precise physical meaning to, the algebra of local observables. (\cite{Bell:1975aa}, p.1)
\end{quote}

\section{Conclusions}
One of the main methodological lessons of the primitive ontology approach is that we should keep as distinct as possible the mathematical, physical and philosophical aspects of a given theory. We should resist the temptation to interpret literally the mathematical structures of physical theories and, as a consequence, one should not infer ontological conclusions directly from their formalisms. It has been one of the main messages of the XX century literature on the foundations of quantum physics that to infer such conclusions from an ontologically ill-defined formalism will lead to questionable positions as the ones we have carefully analyzed in the previous section. The case of Malament's theorem is particularly important for the philosophical foundations of quantum field theory given its importance and the general consensus it has received. To see how Laudisa's critical remarks apply in this case is notable fact, and we are forced to seriously re-think the issue of the ontology of QFT, a matter which is not already settled.

I completely share Laudisa's worries concerning the move to draw ontological conclusions from mathematical results which in turn rely upon metaphysical assumptions which have been seriously questioned, and I hope the reader will be convinced of this fact as well. However, I disagree with Laudisa on Gisin's theorem, a result which should be considered for philosophical reflections within the primitive ontology community, because of its import for the possibility to make a theory as BM compatible with special relativity. 

In conclusion, no-go theorems in quantum physics provide useful information concerning the boundaries of a specific theoretical framework as well as the possible knowledge we might have on physical systems, as stated in \cite{Cowan:2016aa}; nevertheless, we should require that their conclusions must be inferred from an ontologically well-defined theory.
\vspace{10mm}

\textbf{Acknowledgements} I would like to thank Federico Laudisa for his comments on the previous draft of this paper. I am grateful to the Swiss National Science Foundation for financial support (Grant No. 105212-175971).
\clearpage

\bibliographystyle{apalike}
\bibliography{PhDthesis}
\end{document}